\documentclass[epjCONF]{svjour}
\pdfoutput=1
\usepackage{graphicx}
\usepackage[varg]{txfonts} % Times fonts
\usepackage[latin1]{inputenc}
\newcommand {\et}            {\mbox{${E_T}$}}
\newcommand {\gev}           {\mbox{$\,\rm GeV$}}

\newcommand {\gevcsq}        {\mbox{$\,\rm GeV\!/{\it c}^2$}}
\newcommand {\met}           {\mbox{${\not \! E}_{T}$}}
\newcommand {\ppbar}         {\mbox{$p\bar{p}$}}
\newcommand {\pt}            {\mbox{$p_{T}$}}
\newcommand {\zee}           {\mbox{$\rm Z \rightarrow e^+ e^-$}}
\newcommand {\zll}           {\mbox{$\rm Z \rightarrow \ell^+ \ell^-$}}

\newcommand {\znn}           {\mbox{$\rm Z \to \nu \nu$}}
\newcommand {\zmm}           {\mbox{$\rm Z \rightarrow \mu^+ \mu^-$}}
\newcommand {\zzllll}        {\mbox{$\rm ZZ \to \ell^+ \ell^- \ell^+ \ell^-$}}
\newcommand {\zzllnn}        {\mbox{$\rm ZZ \to \ell^+ \ell^- \nu \nu$}}
\newcommand {\zzlljj}        {\mbox{$\rm ZZ \to \ell^+ \ell^- j j$}}
\session-title{XXIInd Hadron Collider Physics Symposium}
\begin{document}
\title{Diboson Physics at the Tevatron}
\author{Aidan Robson\inst{1}\fnmsep\thanks{\email{aidan.robson@glasgow.ac.uk}} for the CDF and D0 Collaborations}
\institute{$^1$ SUPA, School of Physics and Astronomy, University of Glasgow, G12 8QQ, Scotland}
\abstract{ Tevatron diboson measurements are reviewed, and new or recent results reported for W$\gamma$, Z$\gamma$ and ZZ production in the leptonic decay modes, and for WW/WZ production in the lepton plus jets decay mode.  The most stringent limits on anomalous triple gauge couplings are reported for each final state.
}
\maketitle
\section{Introduction}
\label{intro}
Following the ending of the Tevatron collider program, we can 
review the significant progress that has been made in the 
diboson sector over the ten years of Run 2.
At the start of Run 2, WW was the only massive diboson state to 
have been measured, with only a handful of events.  In the 
intervening years, the WZ and ZZ processes have been observed 
(in 2007 and 2008 respectively), and the new availability of 
theoretical tools such as 
{\sc mcfm} \cite{mcfm} and {\sc mc@nlo} \cite{mcnlo} has allowed 
the standard model to be tested in the diboson sector.
Measuring diboson production addresses the basic physics interest 
of observing fundamental electroweak processes.
Measuring increasingly small cross-sections is a stepping-stone 
to new physics; and as diboson production is a major background 
to Higgs searches, it is important to understand it.  Furthermore, 
measuring diboson production allows access to triple gauge 
couplings, which could provide indications of new physics.

\section{W/Z + photon}

\subsection{W$\gamma$}

\begin{figure}[h!]
  \begin{center}
    \includegraphics[width=0.45\textwidth, clip=true, viewport=0.1in 0.1in 8in 6.in] 
    {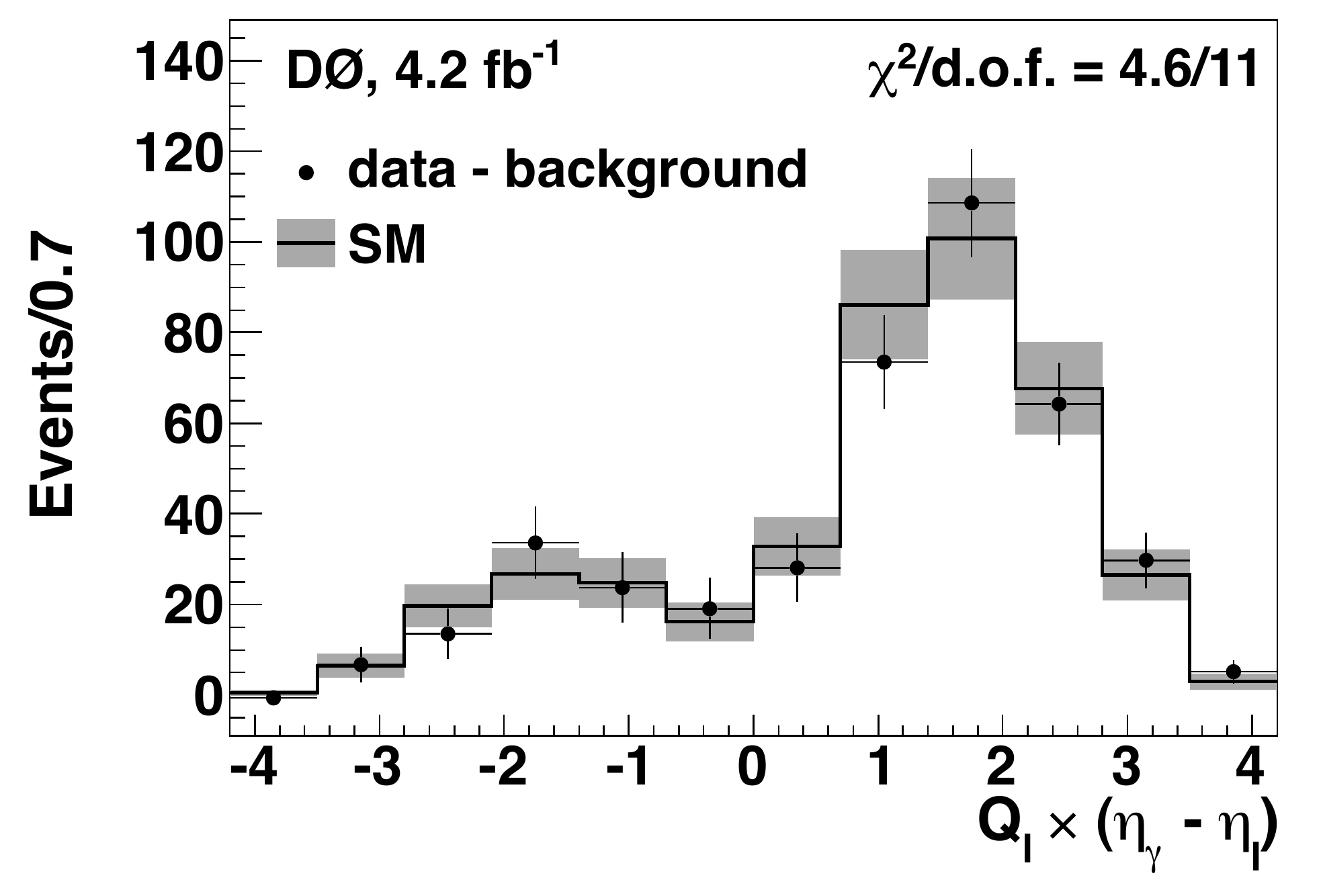}
    \includegraphics[width=0.45\textwidth, clip=true, viewport=0.in -0.3in 7.5in 5.in] 
    {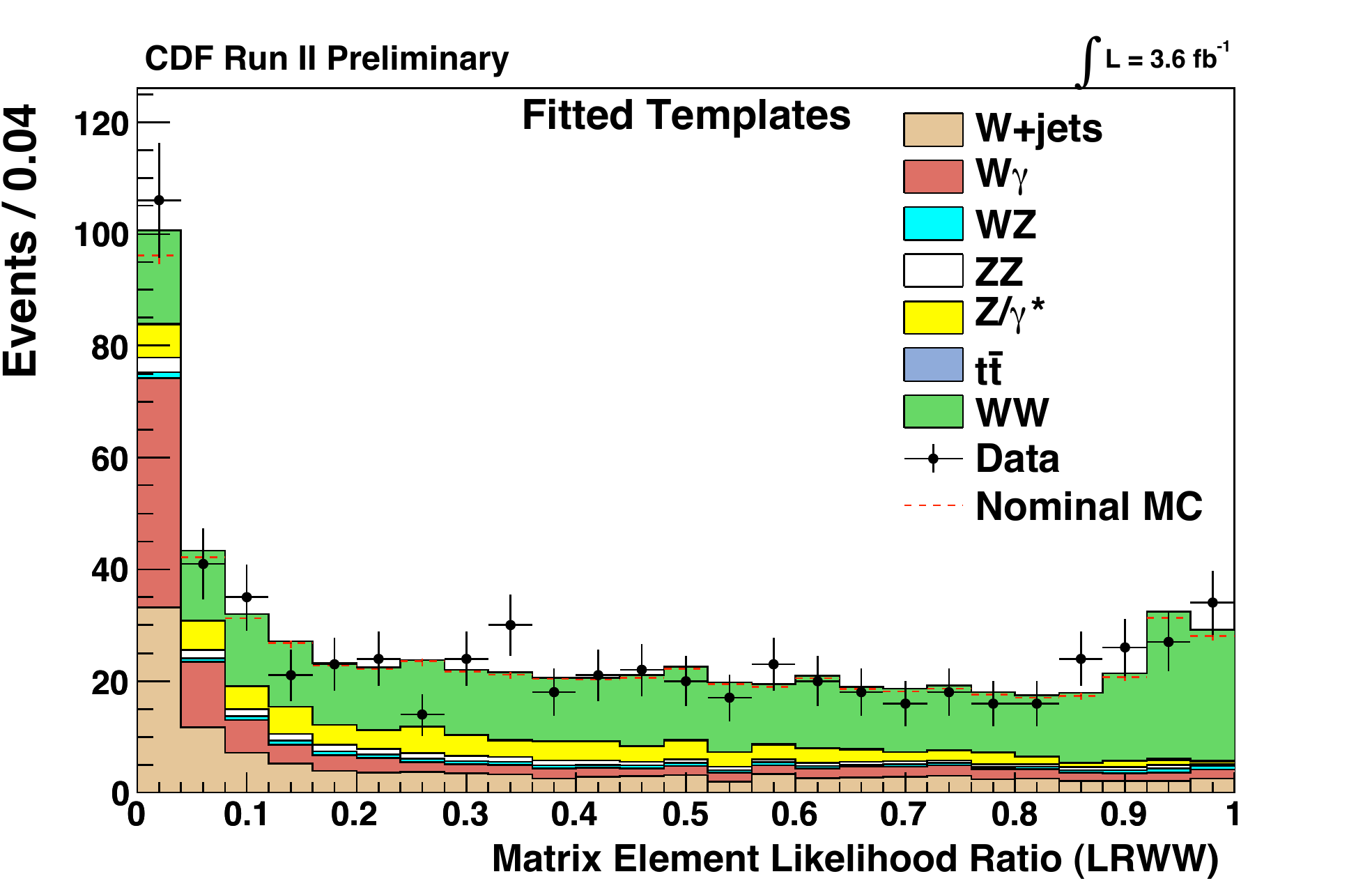}
    \caption[]{ 
	(upper) The charge-signed photon-lepton rapidity difference for W$\gamma$ candidates from D0, showing the radiation amplitude zero as a dip at around -1/3; and (lower) the matrix element likelihood discriminant for WW candidates from CDF.
    }
\vspace*{-.4cm}
    \label{fig:wgamma}
  \end{center}
\end{figure}

D0 has a new result in W$\gamma$ production from September this year, 
using 4.2\,fb$^{-1}$ of integrated luminosity.
Events are selected with an electron or muon, a photon, and missing 
transverse energy (\met).
This analysis uses a neural network for photon identification to 
improve sensitivity to WW$\gamma$ coupling.  Backgrounds are at 
the $20-25\%$ level, overwhelmingly W+jets, and are estimated from data.
An important property of the standard model prediction at leading order
is that interference between the $s$- and $t$-channel amplitudes 
produces a zero in the total W$\gamma$ yield at a specific angle 
$\theta^*$ between the W boson and the incoming quark in the W$\gamma$ 
rest frame.
Although it is difficult to measure the angle directly, this so-called 
`radiation amplitude zero' is also visible in the charge-signed 
photon-lepton rapidity difference as a dip at around -1/3.
Figure~\ref{fig:wgamma} shows the dip, compared with the signal prediction 
from the Baur-Berger dedicated event generator \cite{baur} interfaced to 
{\sc pythia} \cite{pythia} for showering.
The measured cross-section for the kinematic region $\et(\gamma)>15$\gev\ 
and $\Delta R(\ell\gamma)>0.7$ is \\
$\sigma({\rm \ppbar \to W\gamma+Z \to \ell\gamma+X}) =(7.6\pm 0.4{\rm (stat)}\pm 0.6{\rm (sys)})$\,pb, in good agreement with the standard model prediction 
$7.6\pm 0.2$\,pb. 
If there were anomalous triple gauge couplings, the photon \et\ spectrum 
would be modified and more high-\et\ photons observed.
The photon \et\ spectrum may therefore be used to derive limits on 
anomalous WW$\gamma$ couplings.  A binned likelihood fit to data is used, and 
the 1-d limits 95\% CL limits obtained are $-0.4<\Delta\kappa_{\gamma}<0.4$ 
and $-0.08<\lambda_{\gamma}<0.07$ for a new physics scale $\Lambda=2$\,TeV.

\subsection{Z$\gamma$}
\begin{figure}[h!]
  \begin{center}
    \includegraphics[width=0.4\textwidth, clip=true, viewport=0.1in 0.1in 6.5in 7.in] 
    {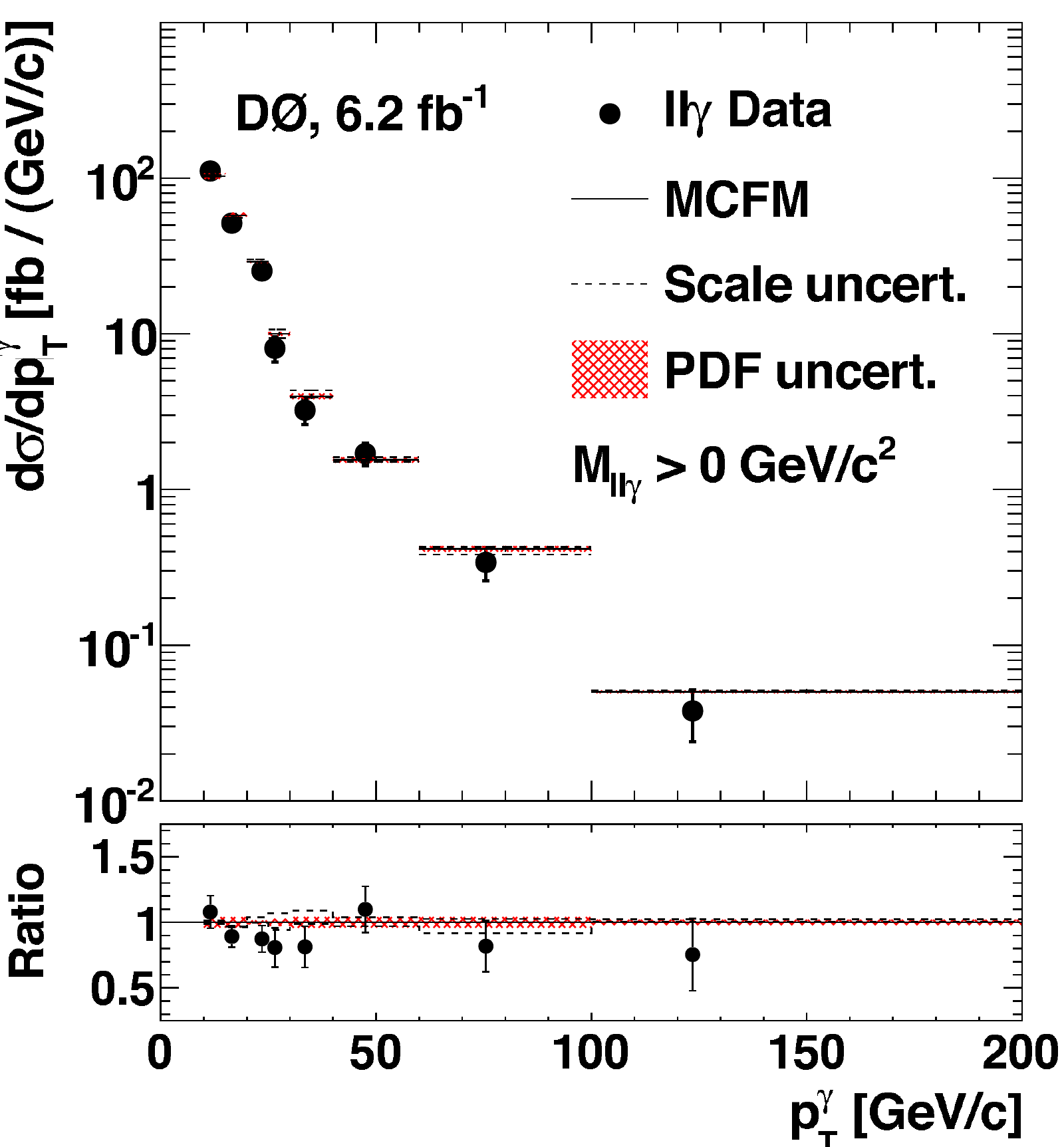}
    \includegraphics[width=0.4\textwidth, clip=true, viewport=0.1in 0.1in 6.5in 7.in] 
    {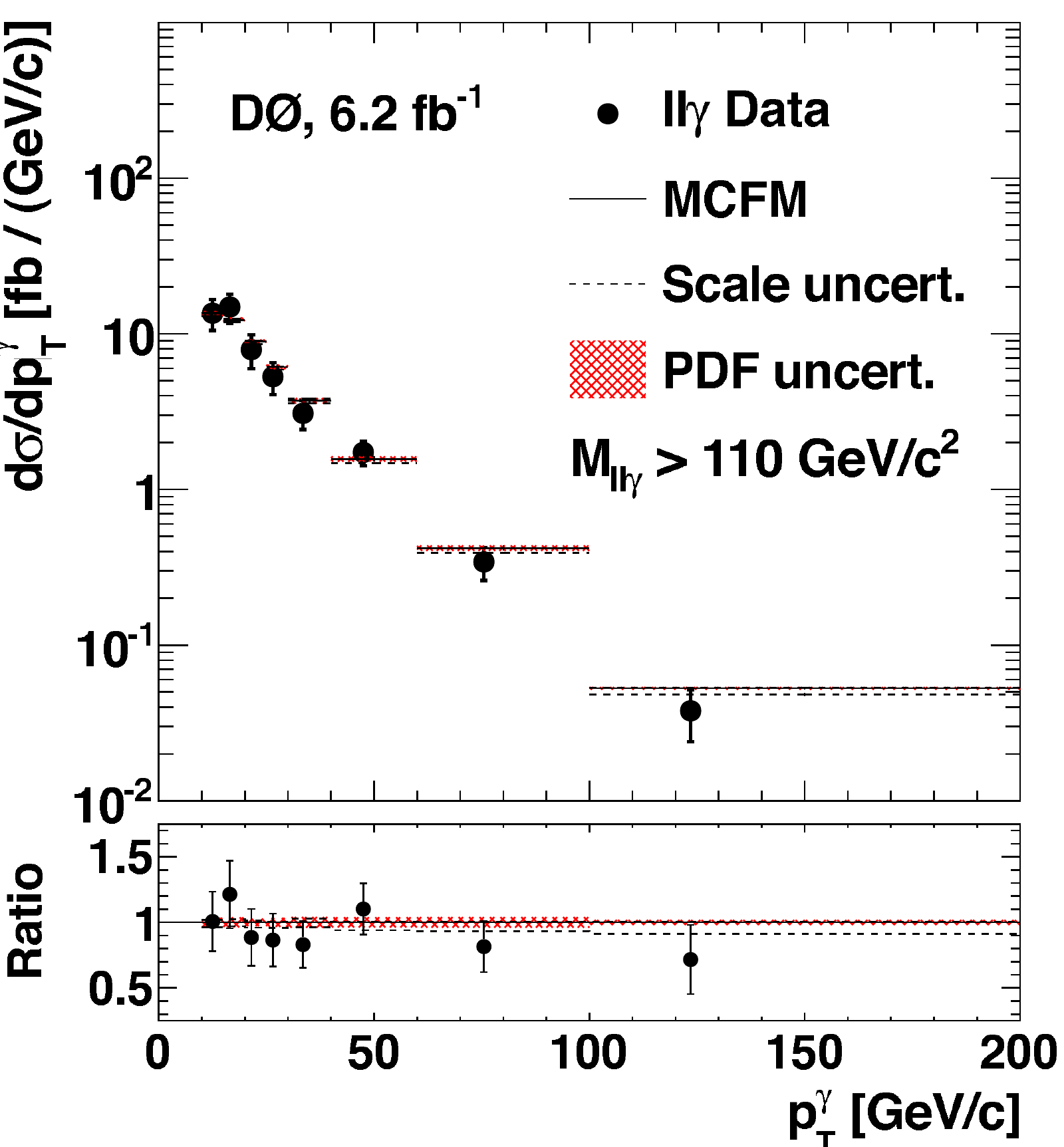}
    \caption[]{ 
	The differential cross-section $d\sigma/dp_T(\gamma)$ for Z$\gamma$ events from D0, for (upper) all values of $M(\ell\ell\gamma)$, and (lower) the ISR-enhanced dataset $M(\ell\ell\gamma)>110$\gevcsq.
    }
    \label{fig:zgamma}
\vspace*{-.4cm}
  \end{center}
\end{figure}

D0 has a new result for this conference in Z$\gamma$, using 
6.2\,fb$^{-1}$ of integrated luminosity.
Again, a neural network technique that uses five variables from 
tracking, calorimetry, and the preshower detectors provides 
robust differentiation between photons and jets.
Background is at the $5-10\%$ level and is dominated by Z+jets.
Around 1000 events are observed in each of the final states 
$\zee+\gamma$ and $\zmm+\gamma$.
The Z$\gamma$ system has the property that initial state photon 
radiation (ISR) may be selected preferentially over final state photon 
radiation by requiring the three-body invariant mass $M(\ell\ell\gamma)$
to be above the Z boson mass.  With $M(\ell\ell\gamma)>110$\gevcsq, 
around 300 events are observed in each of the final states.
The differential cross-section $d\sigma/dp_T(\gamma)$ is measured, 
using matrix inversion to unfold the experimental distribution, 
and is shown in Figure~\ref{fig:zgamma} both for all $M(\ell\ell\gamma)$, 
and for the ISR-dominated sample $M(\ell\ell\gamma)>110$\gevcsq.
Prior to this analysis, these differential distributions had not 
been shown.  The data are compared with the NLO prediction from {\sc mcfm},
and are seen to be consistent.  
Total cross-sections are also quoted: for the kinematic region
$|\eta(\gamma)|<1$, $\et(\gamma)>10$\,GeV, $\Delta R(\ell\gamma)>0.7$ 
and $M(\ell\ell\gamma)>60$\gevcsq\ the result is 
$\sigma({\rm \ppbar \to Z\gamma \to \ell\ell\gamma})=(1.09\pm 0.04{\rm (stat)}\pm 0.07{\rm (sys)})$\,pb, to be compared with the standard model prediction 
$1.10\pm 0.03$\,pb; and for $M(\ell\ell\gamma)>110$\gevcsq\ the 
result is 
$\sigma({\rm \ppbar \to Z\gamma \to \ell\ell\gamma})=(0.29\pm 0.02{\rm (stat)}\pm 0.01{\rm (sys)})$\,pb, to be compared with the standard model prediction 
$0.29\pm 0.01$\,pb.

The most stringent anomalous coupling limits in Z$\gamma$ are from 
CDF, in another 2011 result \cite{cdfzgamma}.  
Here, $\zll+\gamma$ events are selected with $M(\ell\ell\gamma)>100$\gevcsq, 
and \znn\ is also included through events having 
$\met >50$\gev.  For the $\znn+\gamma$ selection, events with tracks
having $\pt >10$\gev\ or jets having $\et>15$\gev\ are rejected, and 
calorimeter timing information is used to reject cosmic-ray tracks.
The photon \et\ spectrum shows no evidence for anomalous couplings 
and is used to set limits; at 95\% CL they are
$-0.020<h_3^Z<0.021$, $-0.0009<h_4^Z<0.0009$, $-0.022<h_3^\gamma<0.020$, 
and $-0.0008<h_4^\gamma<0.0008$, for $\Lambda=1.5$\,TeV.

D0 has also used the Z$\gamma$ signature to look for resonances
and has set limits on generic scalar or vector resonances, such as 
might occur in technicolour models, at the level of 1\,pb \cite{d0zgammares}.

\section{Massive Dibosons}

\subsection{WW}

The WW final state is intimately connected with Higgs searches,
and CDF's WW measurement was done in parallel with the search 
for ${\rm H\to WW}$, using the same tools \cite{cdfww}.  Events 
having two oppositely-charged leptons and \met\ are selected.  
Around 12\% of the acceptance comes from $\tau$ leptons decaying to 
electrons or muons.  Control samples such as same-sign dileptons 
check background modelling. 
The analysis uses a matrix element probability approach, where 
transfer functions derived from simulation are applied to 
the measured four-vectors, which are inputs to matrix elements 
that allow the computation of probabilities that an event comes 
from signal or one of several background processes.  These 
probabilities are put together in a likelihood ratio, and the 
cross-section is extracted from a fit as shown in Figure~\ref{fig:wgamma}.  
With 3.6\,fb$^{-1}$ of integrated luminosity, CDF measures 
$\sigma({\rm \ppbar \to WW})=(12.1\pm 0.9{\rm (stat)} ^{+1.6}_{-1.4} {\rm (sys)})$\,pb.
A small excess in the high tail of the lepton \pt\ spectrum makes  
anomalous triple gauge coupling limits less stringent than 
expected.  The probability that the observed distribution is drawn 
from the standard model is not too small (7\%), so the small excess 
is ascribed to a statistical fluctuation.  The best 
limits come from D0's 1\,fb$^{-1}$ analysis \cite{d0ww}, and at 95\% CL are 
$-0.54<\Delta\kappa_{\gamma}<0.83$,  
$-0.14<\lambda_{\gamma}=\lambda_{Z}<0.18$, and 
$-0.14<\Delta g_1^Z<0.30$
 for a new physics scale $\Lambda=2$\,TeV.

\subsection{WZ}
\begin{figure}[h!]
  \begin{center}
    \includegraphics[width=0.4\textwidth, clip=true, viewport=0.1in 0.in 7.8in 6.9in] 
    {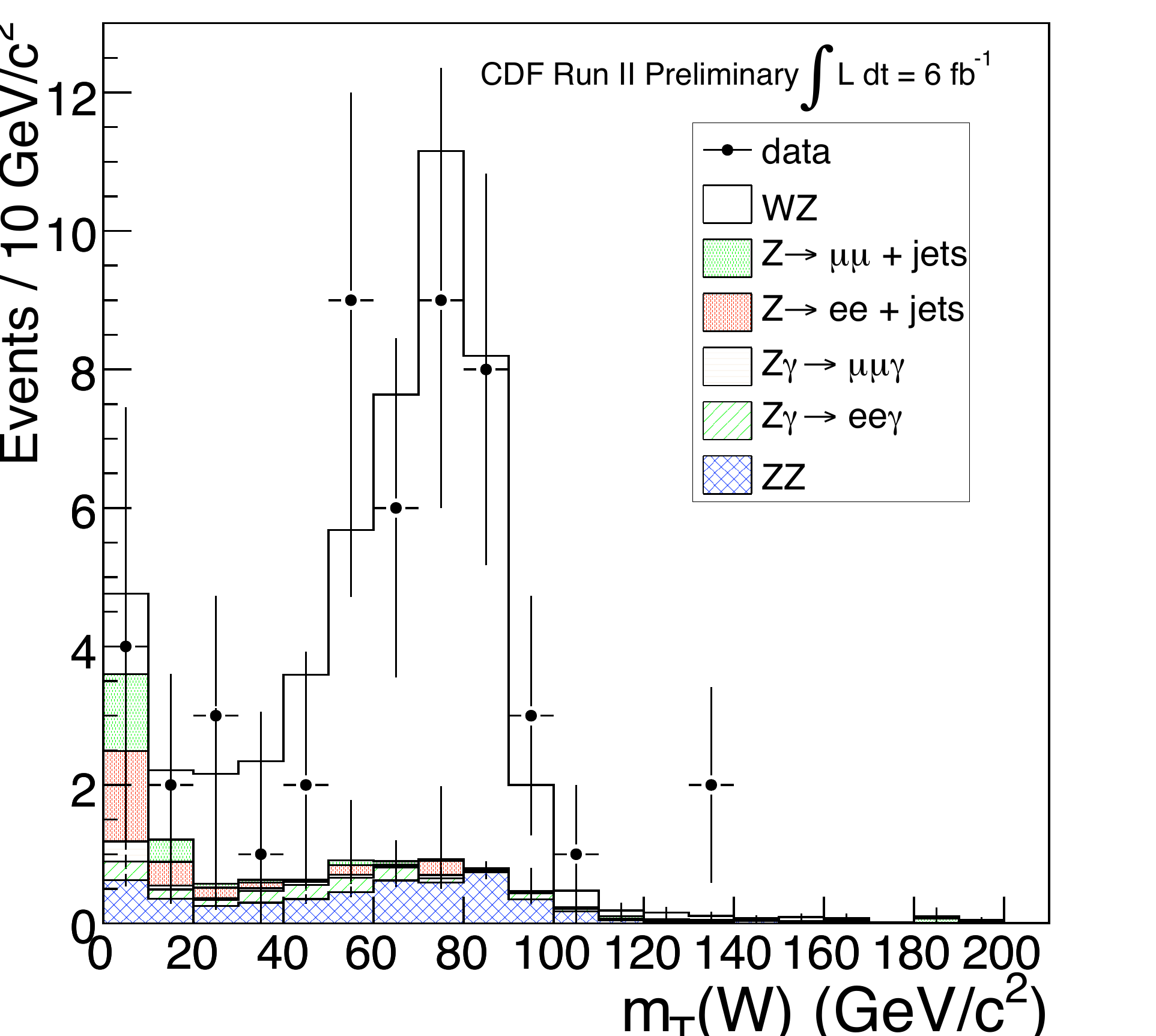}
    \includegraphics[width=0.55\textwidth, clip=true, viewport=0.1in 0.in 8.5in 6.in] 
    {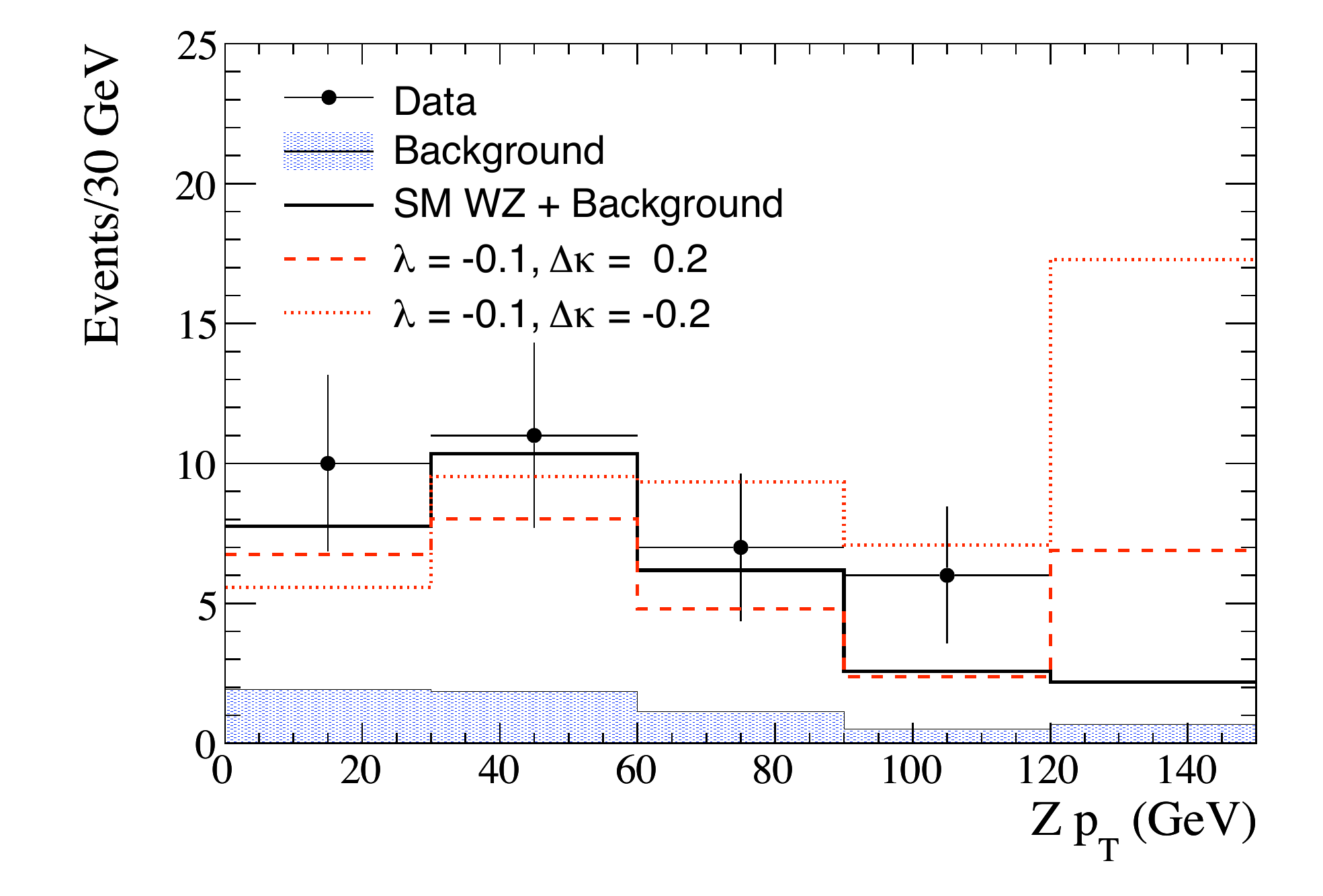}
    \caption[]{ 
      (upper) The W boson transverse mass $m_T$ in WZ candidates recorded by CDF; and (lower) the Z boson \pt\ in WZ candidates recorded by D0.
    }
\vspace*{-.4cm}
    \label{fig:wz}
  \end{center}
\end{figure}

The WZ final state is little-studied as it is charged, and therefore 
produced only at hadron colliders.
CDF's recent analysis in the $\ell\ell\ell\nu$ final state 
using 6\,fb$^{-1}$ of integrated luminosity 
incorporates improvements in lepton selection and shows very good 
resolution, as demonstrated by the W boson transverse mass in Figure~\ref{fig:wz}.
The measured cross-section is normalised to the measured Z boson 
cross-section to remove some systematic uncertainties,
  in particular the luminosity uncertainty.
This is then exchanged for a smaller theoretical 
  uncertainty when multiplying by a calculation of 
  the Z boson production cross-section in order to recover the WZ cross-section:
$\sigma({\rm \ppbar \to WZ})/\sigma({\rm \ppbar \to Z})=(5.5\pm 0.9)\times 10^{-4}$ 
and $\sigma({\rm \ppbar \to WZ})=(4.1\pm 0.7)$\,pb. 
Triple gauge couplings were not studied in this analysis and the best 
anomalous coupling limits are set by D0's 4.1\,fb$^{-1}$ analysis, 
using the Z boson \pt\ distribution shown in Figure~\ref{fig:wz}:
$-0.400<\Delta\kappa_{Z}<0.675$,  
$-0.077<\lambda_{Z}<0.093$, and 
$-0.056<\Delta g_1^Z<0.154$
 for a new physics scale $\Lambda=2$\,TeV \cite{d0wz}.

%D0 has additionally used the WW/WZ signature to search for resonances, 
%and has set limits on W' and RS graviton production \cite{d0wzres}.

\subsection{ZZ}

\begin{figure}[h!]
  \begin{center}
    \includegraphics[width=0.45\textwidth, clip=true, viewport=0.1in 0.in 7.8in 6.5in] 
    {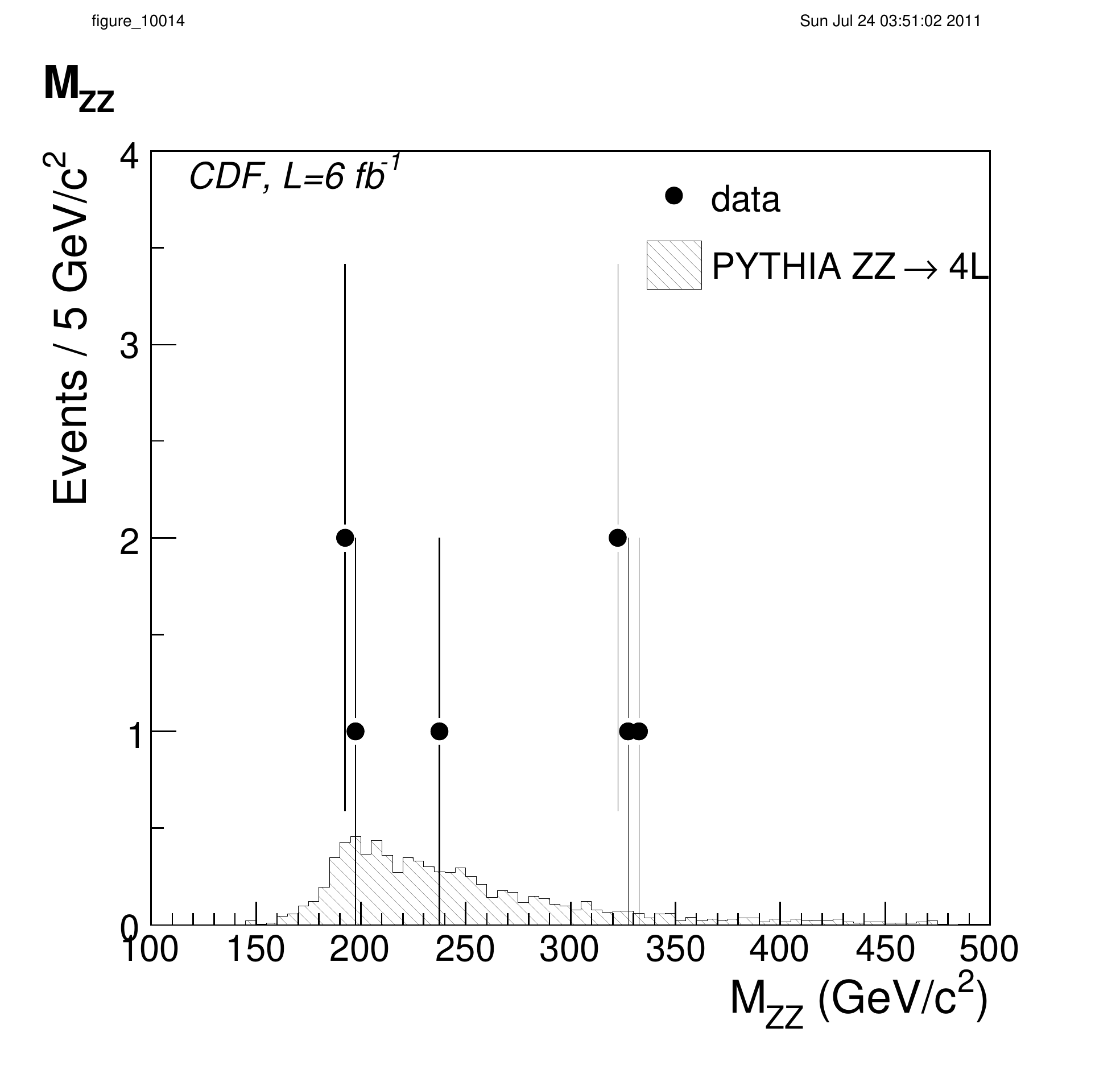}
    \includegraphics[width=0.39\textwidth, clip=true, viewport=0.1in -.2in 8.in 7.8in] 
    {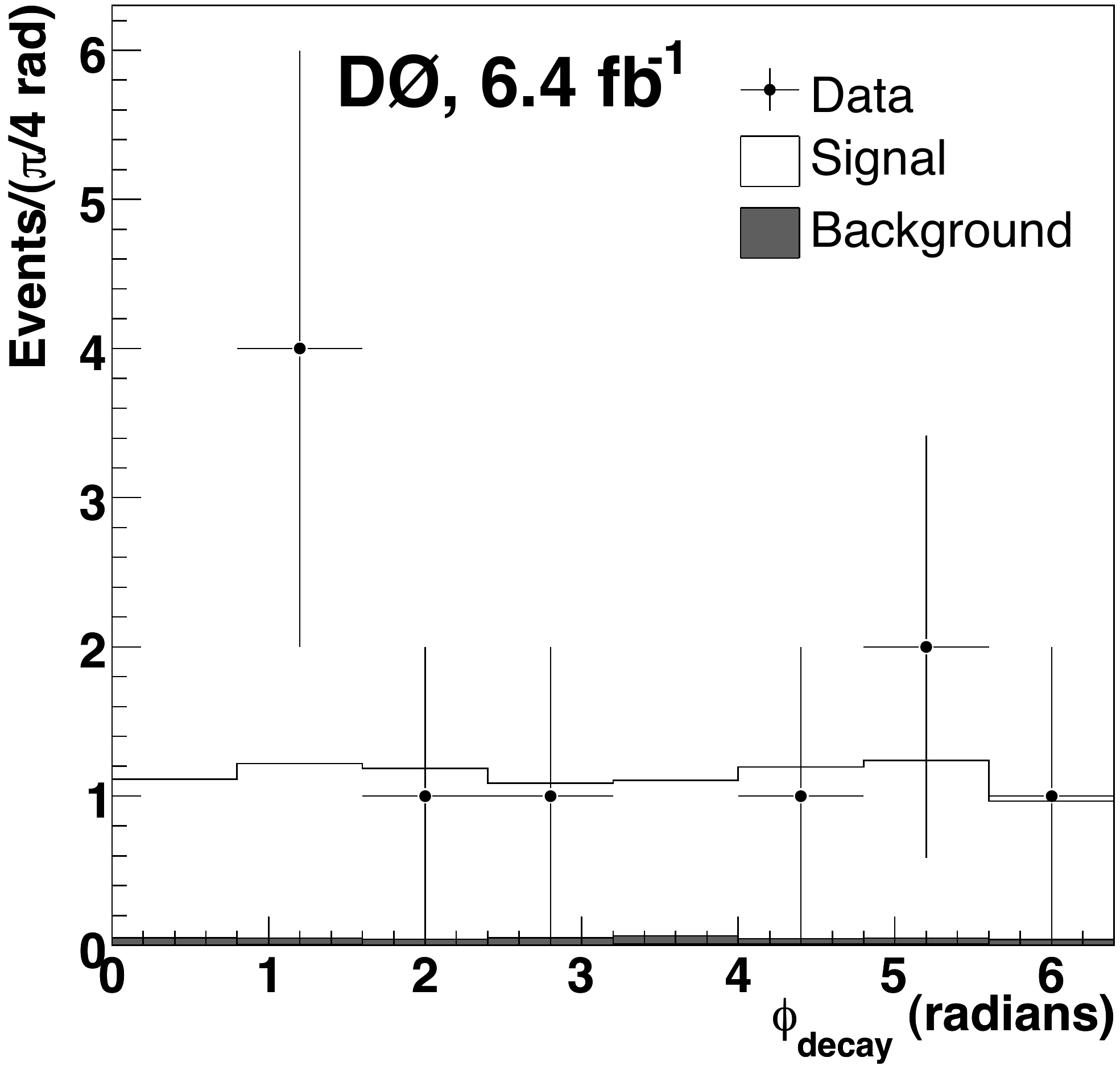}
    \caption[]{ 
      (upper) The reconstructed four-lepton mass in CDF \zzllll\ candidates, and (lower) angular separation of the lepton pair planes, measured in the ZZ rest frame, for D0 \zzllll\ candidates.
    }
\vspace*{-.4cm}
    \label{fig:zz}
  \end{center}
\end{figure}

CDF has a new measurement of the ZZ production cross-section in 
the four-lepton final state using 6\,fb$^{-1}$ of integrated luminosity \cite{cdfzz}: 
$\sigma({\rm \ppbar \to ZZ})=(2.3^{+0.9}_{-0.8}{\rm (stat)}\pm 0.2{\rm (sys)})$\,pb, 
to be compared with the NLO standard model prediction $1.4\pm 0.1$\,pb. 
A clustering of events at high mass, shown in Figure~\ref{fig:zz}, caused 
excitement.  However, analysis of the other ZZ final states 
\zzllnn\ and \zzlljj\ showed them to be more sensitive to a resonance 
of mass around 327\gevcsq\ decaying to ZZ, and the data in those channels 
are in agreement with standard model predictions.  The four-lepton events 
therefore appear to arise from standard model sources.

D0 also has a recent measurement of \zzllll, with increased muon 
acceptance compared to previous results \cite{d0zz}.  The measured cross-section is 
$\sigma({\rm \ppbar \to ZZ})=(1.26^{+0.47}_{-0.37}{\rm (stat)}\pm 0.14{\rm (sys)})$\,pb. 
The distribution in angle between the planes of the lepton pairs, computed 
in the ZZ rest-frame, is sensitive to the production mechanism of the Z
pair: for example, a Z pair arising from the decay of a Higgs boson would 
result in a different angular distribution.  This distribution is tested 
for the first time and is shown in Figure~\ref{fig:zz}; it is seen to be consistent 
with the standard model expectation.

The only anomalous triple coupling limits from ZZ are from an earlier D0 
result \cite{d0zz1fb}:
$-0.28<f_{40}^{Z}<0.28$, 
$-0.26<f_{40}^{\gamma}<0.26$, 
$-0.31<f_{50}^{Z}<0.29$, and 
$-0.30<f_{50}^{\gamma}<0.28$ at 95\% CL for $\Lambda=1.2$\,TeV. 

Finally, CDF has measured \zzllnn\ in 5.9\,fb$^{-1}$ of integrated luminosity
using techniques from the Higgs search, and in that channel found \\
$\sigma({\rm \ppbar \to ZZ})=(1.45^{+0.45}_{-0.42}{\rm (stat)} ^{+0.41}_{-0.30}{\rm (sys)} )$\,pb.

\section{Diboson final states with jets}

Given their similarity to key Higgs boson signatures, there have 
been ongoing efforts to observe diboson production in final states 
with jets.  

\begin{figure}[h!]
  \begin{center}
    \includegraphics[width=0.4\textwidth, clip=true, viewport=0.1in 2.5in 7.8in 9.5in] 
    {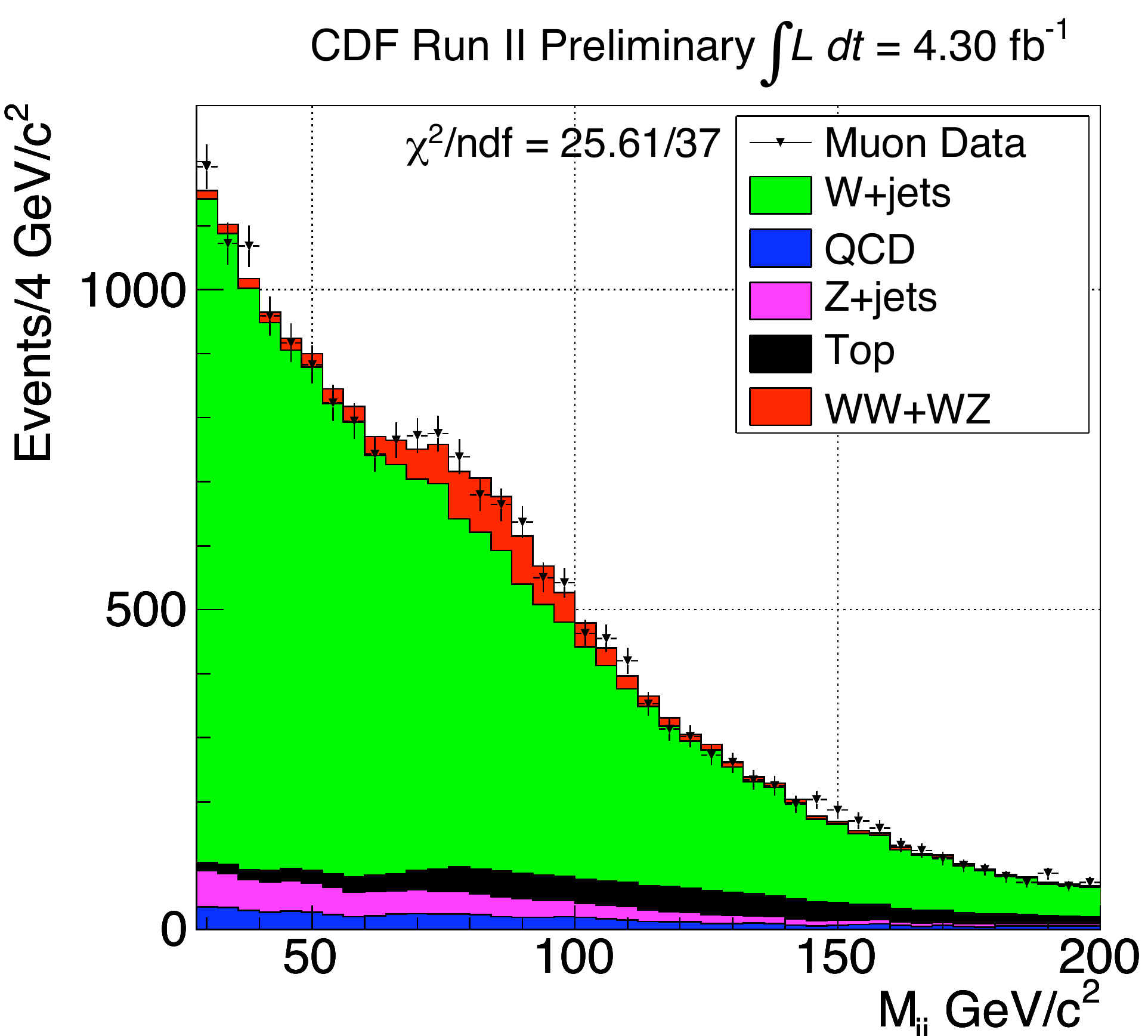}
    \includegraphics[width=0.45\textwidth, clip=true, viewport=0.1in 0.in 7.8in 6.9in] 
    {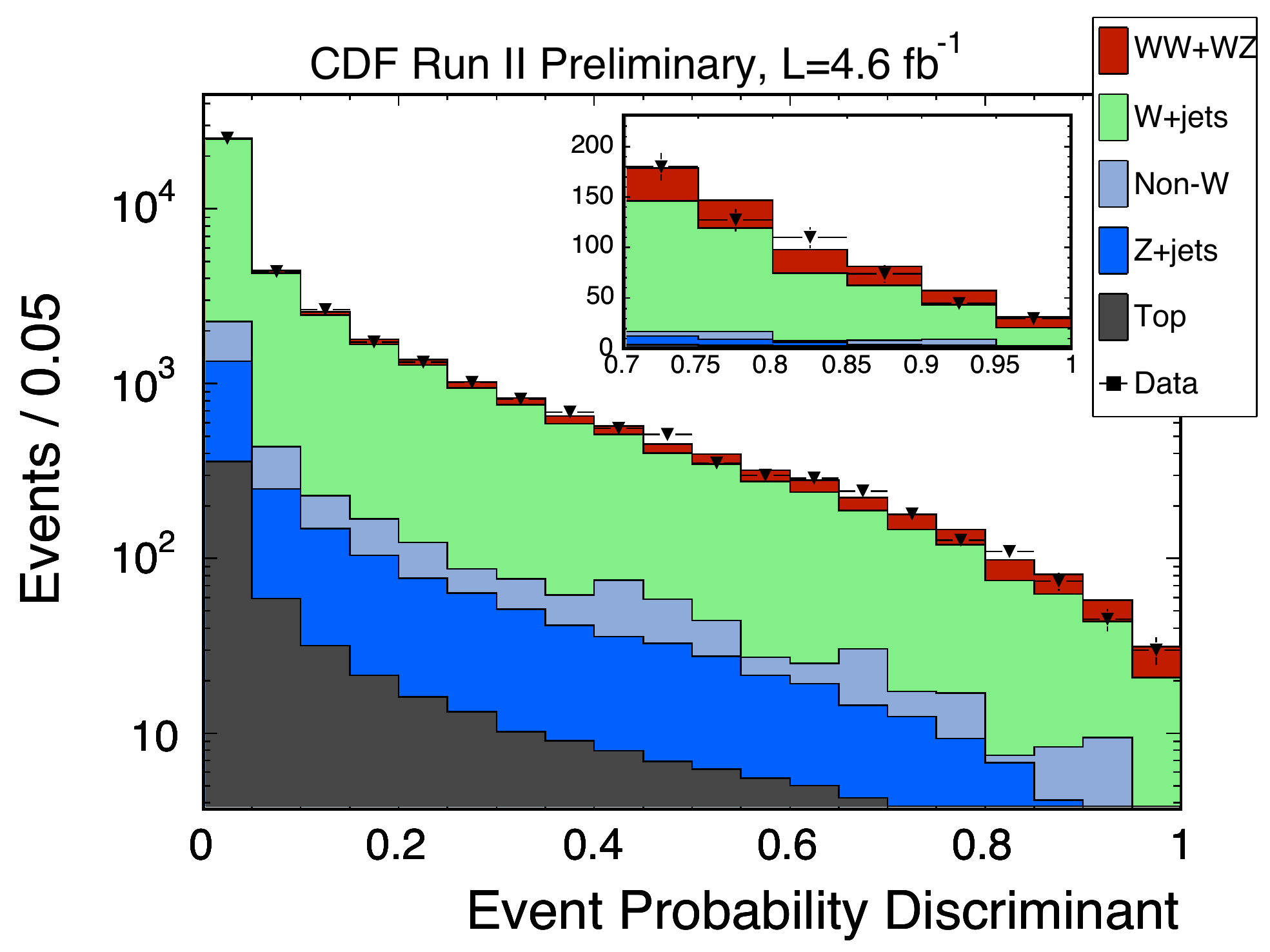}
    \caption[]{ 
	(upper) The dijet invariant mass spectrum in the WW/WZ muon+jets channel from CDF, and (lower) the event probability discriminant in the matrix-element probability approach.
    }
    \label{fig:cdflvjj}
\vspace*{-.4cm}
  \end{center}
\end{figure}

Two CDF analyses observed WW and WZ production in the $\ell\nu jj$ final 
state in 2010.  This final state is very similar to that expected 
from WH associated production.  W+jets is the overwhelming 
background.  In the first analysis, the background contribution from 
QCD was fitted from data using the \met\ distribution, where for 
the analysis selection, QCD enters at low values, and electroweak 
processes have high values.  The signal was extracted from a $\chi^2$ 
fit to the dijet mass distribution as shown in Figure~\ref{fig:cdflvjj}, 
giving an extracted cross-section 
$\sigma({\rm WW+WZ}) = (18.1\pm 3.3 {\rm(stat)}\pm 2.5 {\rm(sys)})$\,pb 
with $5.2\sigma$ significance \cite{cdflvjj_mjj}.
The second analysis used a matrix element technique, for which the 
final event probability discriminant is shown in Figure~\ref{fig:cdflvjj}.
Here, the extracted cross-section was
$\sigma({\rm WW+WZ}) = 16.5^{+3.3}_{-3.0}$\,pb, with $5.4\sigma$ significance \cite{cdflvjj_mjj}.

\begin{figure}[h!]
  \begin{center}
    \includegraphics[width=0.45\textwidth, clip=true, viewport=0.1in 0.in 13.4in 9.5in] 
    {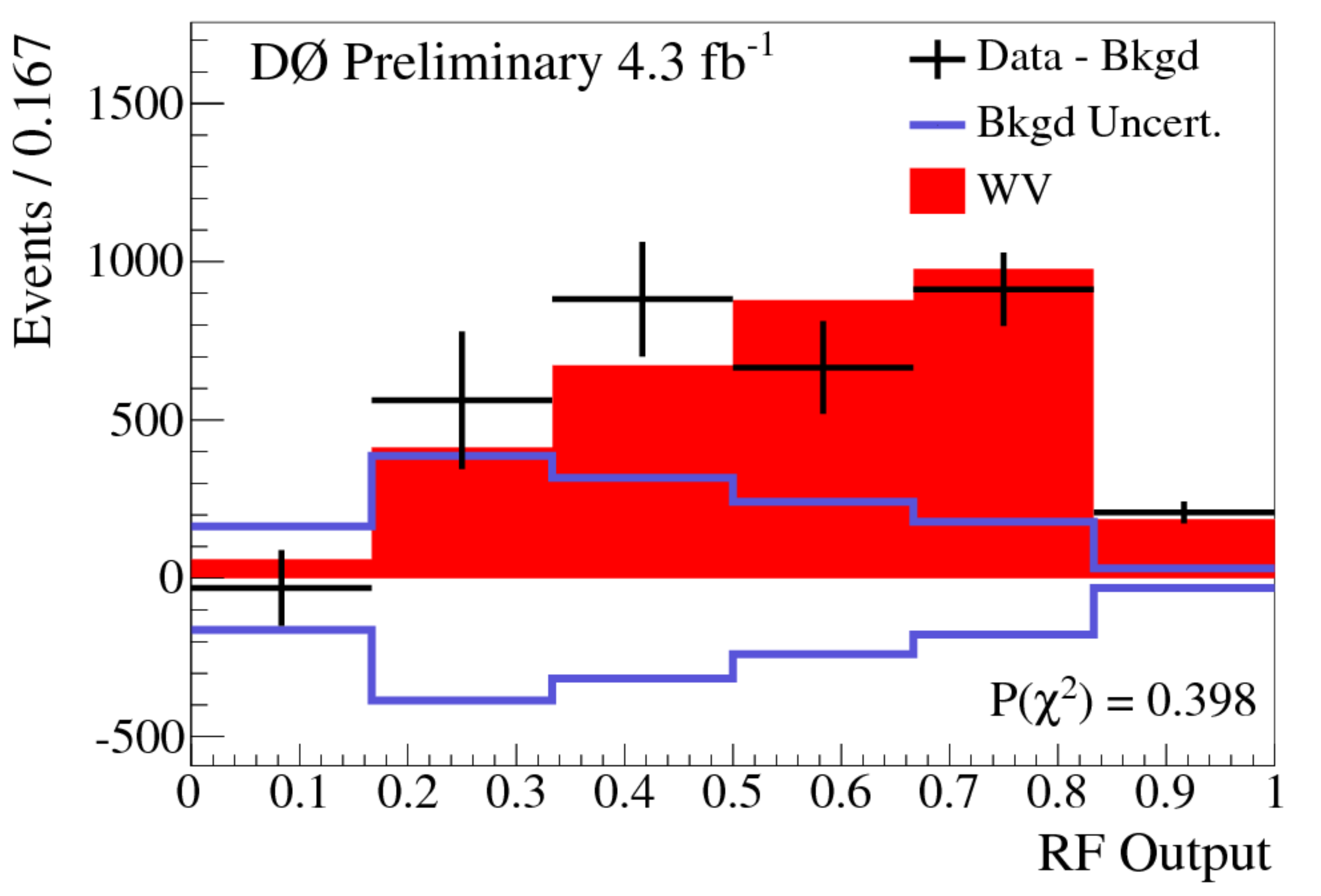}
    \includegraphics[width=0.48\textwidth, clip=true, viewport=0.in 0.1in 13.4in 9.5in] 
    {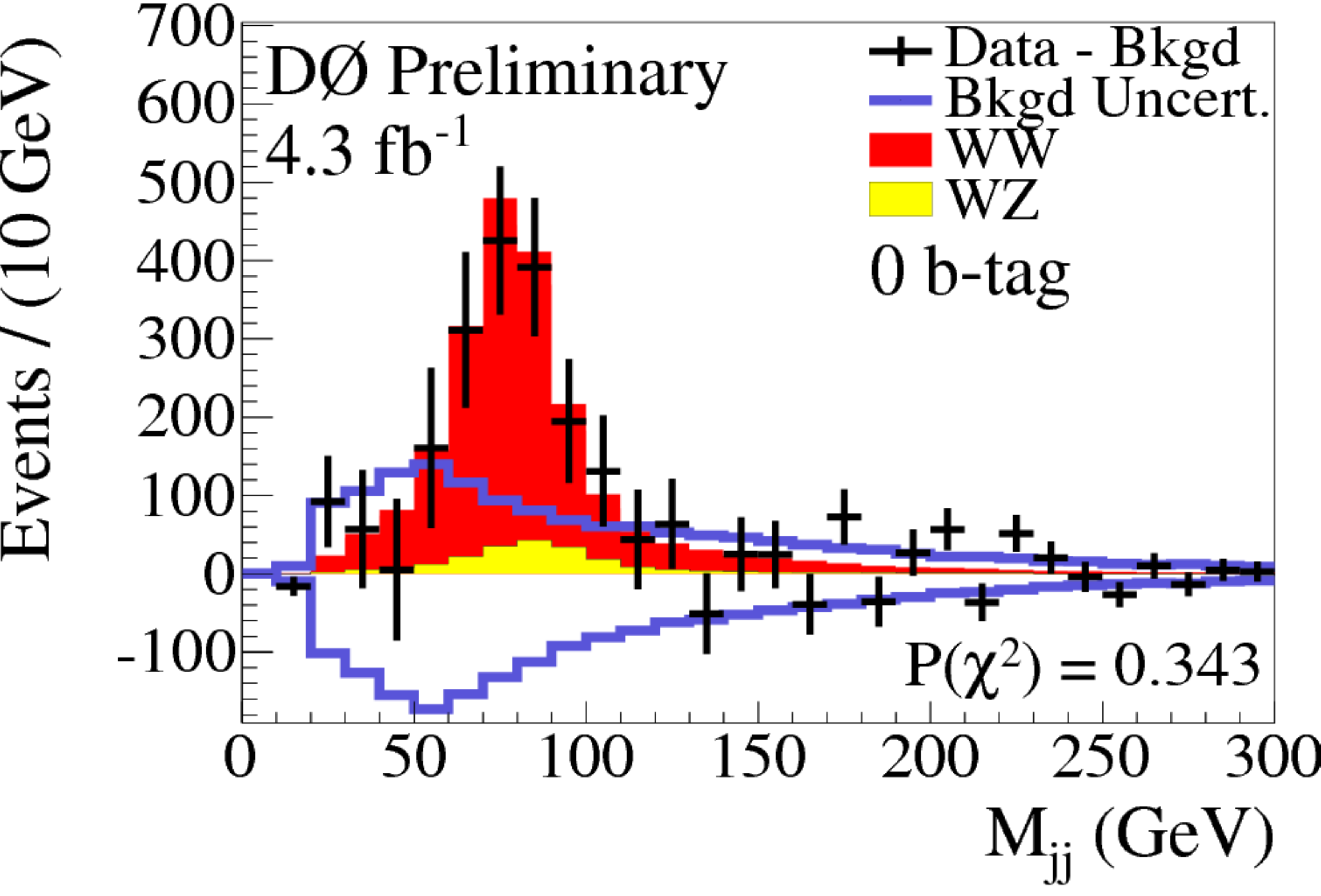}
    \includegraphics[width=0.43\textwidth, clip=true, viewport=0.1in 0.in 13.4in 9.5in] 
    {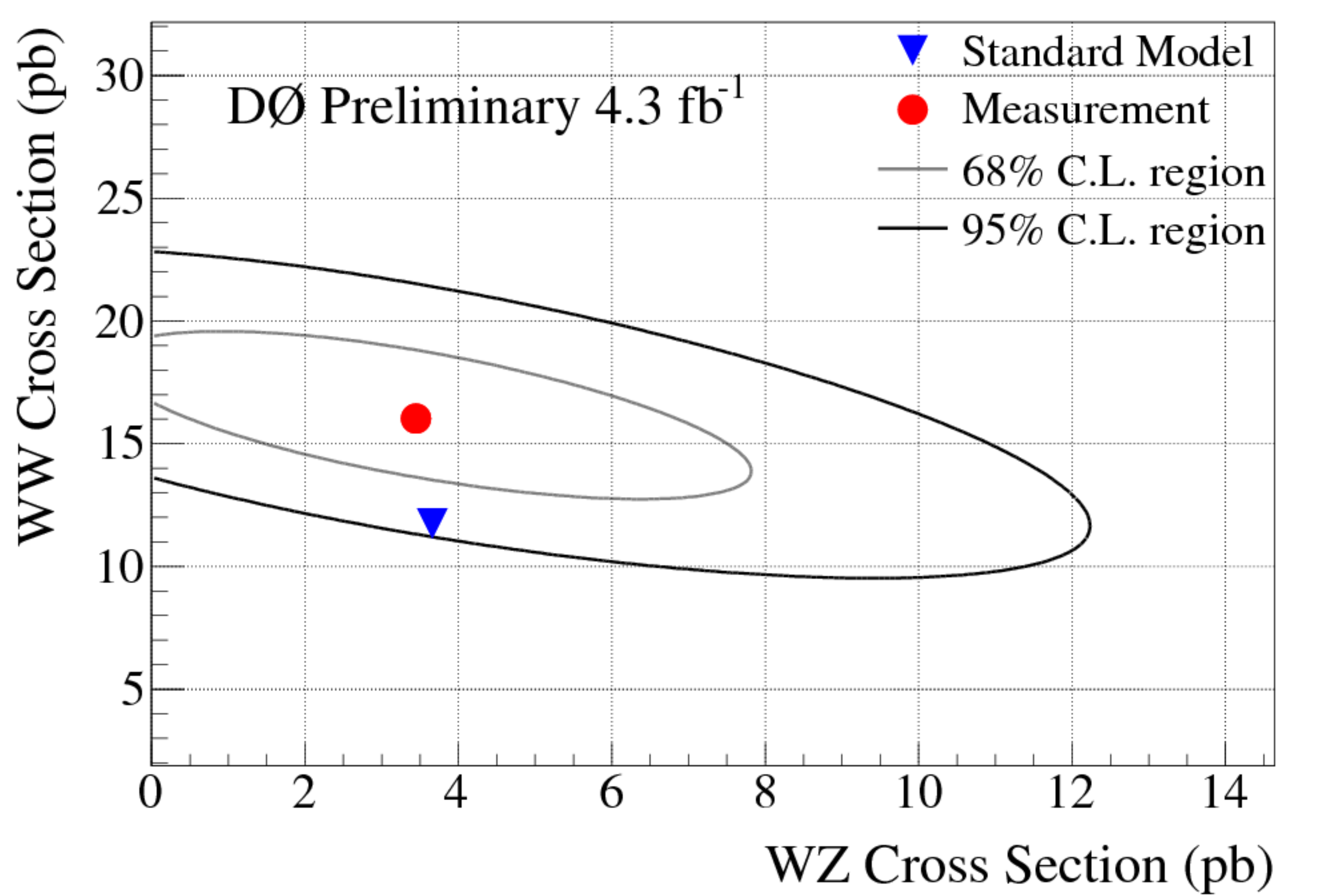}
    \caption[]{ 
	Results from D0's WW/WZ analysis in the $\ell\nu jj$ final state: (upper) random forest multivariate discriminant output; (centre) background-subtracted dijet mass; (lower) contours of WW and WZ production cross-section.
    }
\vspace*{-.4cm}
    \label{fig:d0lvjj}
  \end{center}
\end{figure}

New for this conference is D0's latest update in the $\ell\nu jj$ final state, 
using 4.3\,fb$^{-1}$ of integrated luminosity, which makes significant advances.
A random forest multivariate discriminant is used to separate signal from 
background, and since Z bosons can decay to b-quark pairs but W bosons 
cannot, $b$-tagging is employed both to improve the significance 
of the observation, and to separate the WW and WZ components.
Both the random forest discriminant output, and the dijet invariant 
mass for the no $b$-tag data sample, are shown in Figure~\ref{fig:d0lvjj}.
A cross-section 
$\sigma({\rm WW+WZ}) = 19.6^{+3.1}_{-3.0}$\,pb is measured, with $8\sigma$ significance, 
and contours of the separated WW and WZ cross-sections are given in Figure~\ref{fig:d0lvjj}.

Further results of diboson analyses with decays to b-quark pairs are given 
elsewhere in these proceedings \cite{grivaz}.

An earlier version of this D0 analysis using 1\,fb$^{-1}$ of integrated 
luminosity set limits on anomalous triple couplings:
$-0.44<\Delta\kappa_{\gamma}<0.675$,  
$-0.10<\lambda_{Z}=\lambda_{\gamma}<0.11$, and 
$-0.12<\Delta g_1^Z<0.20$
 at 95\% CL for $\Lambda=2$\,TeV \cite{d0oldwwwzlvjj}.

Finally, an early analysis from CDF in \zzlljj\ using 1.9\,fb$^{-1}$ 
of integrated luminosity set anomalous coupling limits: at 95\% CL, 
$-0.12<f_{4}^{Z}<0.12$, 
$-0.10<f_{4}^{\gamma}<0.10$, 
$-0.13<f_{5}^{Z}<0.12$, and 
$-0.11<f_{5}^{\gamma}<0.11$ for $\Lambda=1.2$\,TeV. 

\section{Outlook}
A rich programme of Tevatron diboson physics has made huge advances 
over the ten years of Run 2, testing the standard model, probing for 
new physics, and underpinning electroweak symmetry-breaking searches.
D0 combined anomalous coupling limits with 1\,fb$^{-1}$ of integrated 
luminosity, resulting in more stringent limits.  Some of those 
have now been superseded, and there is work on a new combination.
Both experiments have a final dataset of around 10\,fb$^{-1}$, so 
as well as being combined, these analyses should be 
updated once more for legacy measurements.

%%\begin{figure}
%%% Use the relevant command for your figure-insertion program
%%% to insert the figure file.
%%% For example, with the option graphics use
%%%\resizebox{0.75\columnwidth}{!}{%
%%%  \includegraphics{fig1.eps} }
%%\caption{Please write your figure caption here.}
%%\label{fig:1}       % Give a unique label
%%\end{figure}
%%%
%%% For tables use
%%\begin{table}
%%\caption{Please write your table caption here.}
%%\label{tab:1}       % Give a unique label
%%% For LaTeX tables use
%%\begin{tabular}{lll}
%%\hline\noalign{\smallskip}
%%first & second & third  \\
%%\noalign{\smallskip}\hline\noalign{\smallskip}
%%number & number & number \\
%%number & number & number \\
%%\noalign{\smallskip}\hline
%%\end{tabular}
%%\end{table}
%%%
%%\begin{thebibliography}{}
%%% and use \bibitem to create references.
%%\bibitem{RefJ}
%%% Format for Journal Reference
%%Author, Journal \textbf{Volume}, (year) page numbers
%%% Format for books
%%\bibitem{RefB}
%%Author, \textit{Book title} (Publisher, place year) page numbers
%%% etc
%%\end{thebibliography}

\end{document}